\documentclass[prl,preprint,superscriptaddress]{revtex4-2}
\usepackage{dcolumn}
\usepackage[mathlines]{lineno}
\usepackage{mathtools}
\usepackage{amssymb}
\usepackage{bm}
\usepackage{pifont}
\usepackage{psfrag}
\usepackage{epstopdf}
\usepackage{wasysym}
\usepackage{amsmath}
\usepackage{tipa}
\usepackage{hyperref}
\usepackage{lineno}
\usepackage{lipsum}
\usepackage[normalem]{ulem}
\usepackage{graphicx}
\usepackage{ragged2e}
\usepackage{caption}
\usepackage{setspace}
\hypersetup{
     colorlinks = true,
     linkcolor = blue,
     anchorcolor = blue,
     citecolor = blue,
     filecolor = blue,
     urlcolor = blue
     }
\usepackage[usenames]{color}
\renewcommand{\footnoterule}{%
  \hrule width \textwidth height 1pt
  \kern 2pt
}

\begin{document}

\title{Reentrant melting of scarred odd crystals by self-shear}
\author{Uttam Tiwari}
\affiliation{Chemistry and Physics of Materials Unit, Jawaharlal Nehru Centre for Advanced Scientific Research, Jakkur, Bangalore - 560064, INDIA}
\author{Pragya Arora}
\affiliation{Chemistry and Physics of Materials Unit, Jawaharlal Nehru Centre for Advanced Scientific Research, Jakkur, Bangalore - 560064, INDIA}
\author{A K Sood}
\affiliation{Department of Physics, Indian Institute of Science, Bangalore - 560012, INDIA}
\affiliation{International Centre for Materials Science, Jawaharlal Nehru Centre for Advanced Scientific Research, Jakkur, Bangalore - 560064, INDIA}
\author{Sriram Ramaswamy}
\affiliation{Department of Physics, Indian Institute of Science, Bangalore - 560012, INDIA}
\affiliation{International Centre for Theoretical Sciences, Bangalore - 560089, India}
\author{Rituparno Mandal}
\affiliation{Soft Condensed Matter Group, Raman Research Institute, Bangalore - 560080, INDIA }
\author{Rajesh Ganapathy}
\affiliation{Chemistry and Physics of Materials Unit, Jawaharlal Nehru Centre for Advanced Scientific Research, Jakkur, Bangalore - 560064, INDIA}
\affiliation{International Centre for Materials Science, Jawaharlal Nehru Centre for Advanced Scientific Research, Jakkur, Bangalore - 560064, INDIA}
\affiliation{School of Advanced Materials (SAMat), Jawaharlal Nehru Centre for Advanced Scientific Research, Jakkur, Bangalore - 560064, INDIA}

\email{Contact author: tiwari@jncasr.ac.in}
\email{Contact author: rajeshg@jncasr.ac.in}

\date{\today}
\begin{abstract}

\textbf{Spatial confinement can induce geometrical frustration in condensed phases, giving rise to topological defects that confer materials with new and exotic properties. Here, we experimentally uncover the remarkable effect of confinement-induced defect strings termed `grain boundary scars' on the behavior of dense two-dimensional assemblies of granular spinners, a canonical odd elastic solid. We show that the spatial arrangement of these scars fundamentally reshapes the flows triggered by chiral activity. Specifically, they cause the topologically protected edge flows - a ubiquitous feature of confined spinner assemblies - to decouple from the bulk. Strikingly, increasing the net chiral activity of the system by tuning the ratio of counterclockwise to clockwise spinners caused spontaneous self-shearing. The resulting odd radial stresses led to a chiral activity-mediated reentrant melting transition at a fixed areal spinner density.  Our findings open new avenues for exploiting geometrical frustration to elicit novel responses from odd elastic solids.}

\end{abstract}
\maketitle

Topological defects are often the orchestrators of material behavior \cite{nelson2002defects}. While their established role lies in tuning the physical properties of passive materials \cite{chaikin1995principles, jangizehi2020defects}, there is burgeoning interest in extending this paradigm to active matter systems \cite{shankar2022topological, ardavseva2022topological}. A simple route to create these defects is through geometrical frustration \cite{sadoc1999geometrical}, achieved by confining the system within a carefully chosen boundary shape that prevents the preferred local order of the system from seamlessly tiling space. These defects cannot be annealed away and remain in the system even in its ground state  - a consequence of Euler's theorem \cite{bowick2009two}. Consider, for instance, the appearance of twelve pentagonal disclination defects in a triangular lattice on a spherical surface; examples of which include viral capsids \cite{bruinsma2021physics}, colloidosomes \cite{bausch2003grain, guerra2018freezing, singh2022observation}, and the familiar soccer ball. Six such defects emerge on confining this lattice within a two-dimensional (2D) circular boundary \cite{moore1999absence, amore2023thomson}. In fact, to relieve the elastic strain energy on exceeding a threshold system size, these disclinations bind to dislocations forming 1D defect strings called `grain boundary scars' \cite{bowick2000interacting, bausch2003grain, guerra2018freezing, singh2022observation, lipowsky2005direct}, which govern the yielding behavior of these frustrated crystals \cite{negri2015deformation}. In the realm of active matter, geometrical frustration effects are more dramatic. In these out-of-equilibrium systems, particularly in those made of elongated building blocks, defects themselves can be motile entities \cite{narayan2007long, marchetti2013hydrodynamics, giomi2014defect, grossmann2020particle, arora2022motile}. This has spurred a surge of research exploring how boundary geometry can be harnessed to create and manipulate these defects, ultimately enabling global control over active flows and turbulence \cite{wu2017transition, memarian2024controlling,liu2021viscoelastic,williams2022confinement, theillard2017geometric, negi2023geometry, hardouin2019reconfigurable, peng2016command, zhang2021spatiotemporal, gompper20202020, ramaswamy2010mechanics, keber2014topology, sknepnek2015active, wioland2013confinement}.

Inspired by these studies, here we examined how confinement-induced topological defects alter the collective behavior of 2D dense assemblies of active spinners. By breaking both parity and time-reversal symmetries, this system allows the existence of novel odd material moduli that are forbidden in equilibrium \cite{fruchart2023odd, tsai2005chiral}. These odd moduli generate transverse couplings between mechanical perturbations and responses; for example, compressing the assembly can induce a net torque density, driving spontaneous rotation. This striking behavior is because interactions between the spinning elements involve transverse forces arising from contact or hydrodynamic friction \cite{banerjee2017odd, mecke2023simultaneous}. While spinners offer a specific realization of oddness, odd mechanics is a more general consequence of chirality and broken time-reversal \cite{kole2024chirality, kole2021layered}. The presence of odd moduli leads to a plethora of new phenomena: topologically protected unidirectional edge currents capable of transporting cargo without scattering \cite{yang2021topologically}, self-sustained chiral waves \cite{tan2022odd}, spontaneous rotation and self-kneading of spinner crystals into whorls \cite{bililign2022motile}, and chiral excitation and tilting of force chains \cite{huang2023odd}. Although many previous studies have investigated confined spinner packings \cite{van2016spatiotemporal, liu2020oscillating, yang2020robust}, the fundamental role of defects stemming from geometrical frustration on the emergent dynamics remains unaddressed, highlighting a key gap in our understanding. 

\begin{figure}[tbp]
\includegraphics[width=1\textwidth]{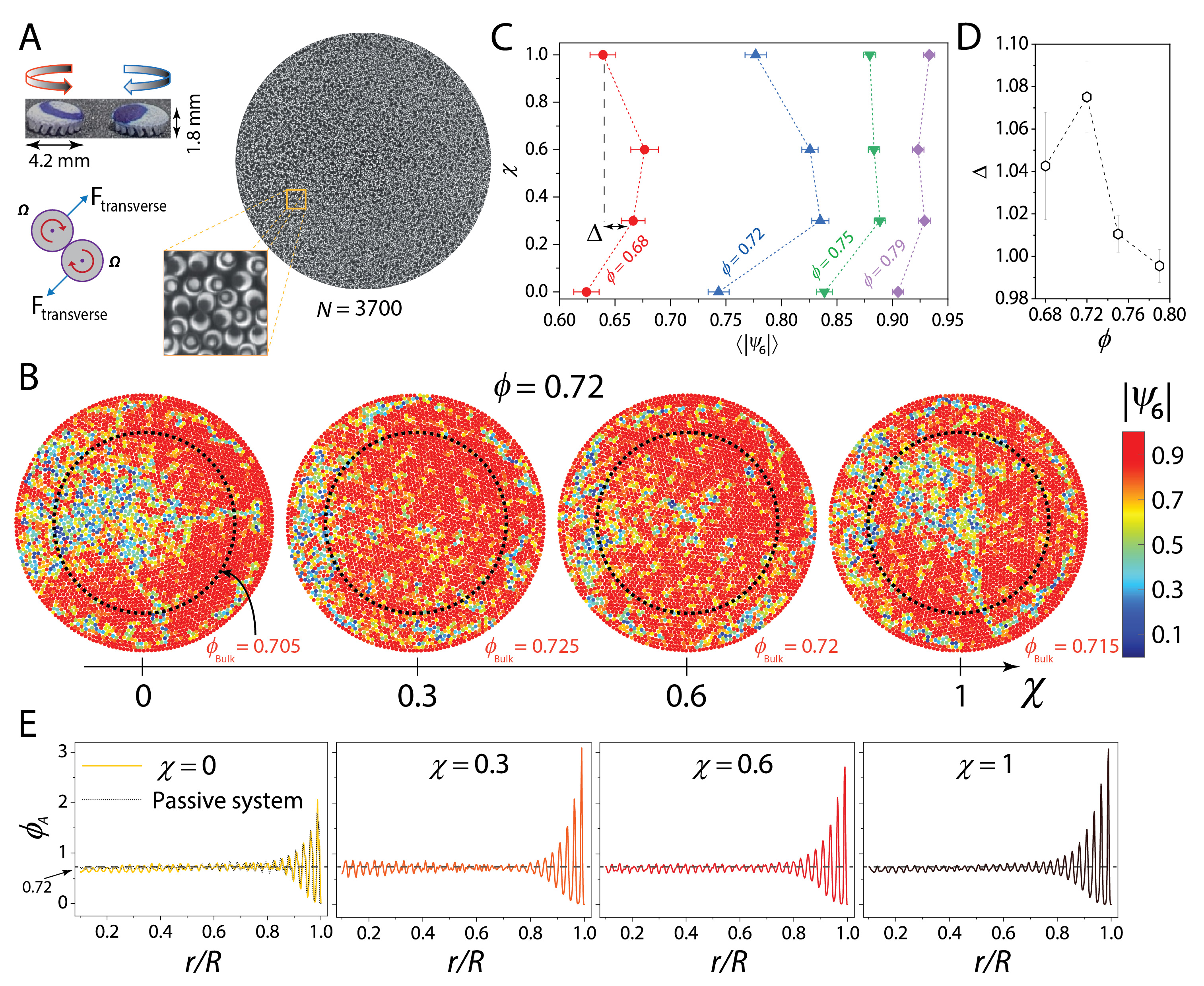} 
\caption{\textbf{Chiral activity drives reentrant melting of spinner crystals.} (\textbf{A}) Top left panel: Picture of 3D-printed clockwise  $(\otimes)$ and counterclockwise $(\odot)$ rotating spinner; arrows show the spin direction. Right panel: Representative image of a dense packing of spinners, interacting via transverse forces, for net chiral activity, $\chi = 0$, and area fraction, $\phi = 0.72$. The zoomed-in view shows circles and dots marked on the $\odot$ and $\otimes$ spinners, respectively, to track their rotation. (\textbf{B}) The panels show spinners colored as per the magnitude of the hexagonal bond-order parameter, $\langle |\psi_6| \rangle$, for $\phi = 0.72$ at different $\chi$ values. The dashed circle delineates the bulk from the boundary and has a cutoff radius, $r_c = 0.7R$, where $R$ is the system's radius. This cutoff corresponds to an $r$ where layering due to confinement is negligible (Fig. \ref{Figure3}C). The averaged areal spinner density of the bulk, $\phi_\text{Bulk}$, is also indicated. (\textbf{C}) shows the variation in $\langle|\psi_6|\rangle$ on increasing $\chi$ for different $\phi$ values. Here, $\langle\rangle$ denotes an average over all the spinners in the bulk and at all times. The error bars denote the standard error. (\textbf{D}) Strength of reentrance, $\Delta$ (shown in Fig. \ref{Figure1}C), versus $\phi$. (\textbf{E}) shows the annular radial density profile, $\phi_A(r)$, with $r/R$, for different $\chi$ values at $\phi = 0.72$. $\phi_A(r)$ for a packing of passive disks at the same $\phi$ is also shown. Here, $r$ is the outer radius of the annulus, with width $0.09d_p$, where $d_p$ is the particle diameter.
\justifying
}
\label{Figure1}
\end{figure}

To single out how confinement-induced topological defects affect the physics of spinner ensembles, it is essential that, sans confinement, the system can attain a defect-free ground state. A second highly desirable feature is to be able to tune the strength of the odd mechanical response that can be elicited out of the material. Granular spinners driven by vertical vibration suit our purpose \cite{scholz2018rotating, petroff2023density} (Fig. \ref{Figure1}A, see Methods). Our spinners are spherical domes of diameter $a = 4.2$ mm standing on tilted legs that transduce vertical vibrations into torque and make them spin. The handedness of rotation, i.e., clockwise $(\otimes)$ or counterclockwise $(\odot)$ is set by the handedness of the leg tilt. Pair-wise spinner interactions comprise a longitudinal hard-core repulsion and a transverse force due to contact friction. The horizontal cross-section of the spinners is circular, and they can readily pack into defect-free triangular lattices. We induced geometrical frustration in the system by confining the spinners within a circular arena of diameter $D = 30$ cm. Our experiments spanned area fractions $0.68\leq\phi\leq0.79$, with the lower limit corresponding to the density beyond which the spinners began to form crystalline domains with triangular symmetry. The internal torque density of the spinner assembly sets the strength of the odd response elicited out of the system \cite{ fruchart2023odd,banerjee2017odd,bililign2022motile}, which we tuned by adjusting the net chiral activity, $\chi$, of the spinner assembly. Transitioning from a racemic mixture (a 50:50 combination of $\otimes$ and $\odot$ spinners) to a homochiral state (all $\otimes$ or all $\odot$) tunes the internal torque density from zero to the maximum achievable while keeping all other parameters fixed.  Here $\chi = {\frac{N^{\otimes} - N^{\odot}}{N}}$ with $N^\otimes$ and $N^\odot$ being the number of $\otimes$ and $\odot$ spinners, respectively, and $N = N^{\otimes} + N^{\odot}$. Fig. \ref{Figure1}A shows a snapshot of the system for $\phi = 0.72$ and $\chi = 0$. For $\phi = 0.79$, $N = 4000$ spinners, making our system size an order of magnitude larger than previous granular experiments \cite{petroff2023density, yang2020robust, scholz2018rotating, liu2020oscillating}.

Fig. \ref{Figure1}B captures our main observation. The panels show spinner packings with increasing $\chi$ for $\phi =0.72$, and the colors represent the magnitude of the hexagonal bond-order parameter, $|\psi^i_6|$, of each spinner. Here, $\psi^i_6 = \frac{1}{N_i} \sum\limits_{j=1}^{N_i} e^{6i\theta_{ij}(t)}$ where, $N_i$ is the coordination of spinner $i$ and $\theta_{ij}$ is the angle made by the line joining the centers of spinner $i$ and its nearest-neighbor $j$ with respect to an arbitrary reference axis. What is striking is the emergence of a near-perfect \textit{single crystal} in the bulk (region enclosed by the dashed circle) for intermediate values of $\chi$, while for the extreme values, a large pocket of liquid is evident. Tuning $\chi$ causes reentrant melting of the bulk, the imprint of which is also present in the pair-correlation function, $g(r)$ (Supplementary Fig. 1). We observe that for equilibrium hard disks in 2D, the liquid phase is found at densities $\phi^\text{Liq}_\text{Eq} \leq 0.716$, and the crystalline phase is stable for $\phi^\text{Xtal}_\text{Eq} \geq 0.72$; the intervening phase is a hexatic \cite{bernard2011two}. While drawing direct comparisons with equilibrium systems is difficult because our system is driven, it is notable that the bulk areal density, $\phi_\text{Bulk}<\phi^\text{Liq}_\text{Eq}$ for $\chi \in[0,1]$ and $\phi_\text{Bulk}\geq\phi^\text{Xtal}_\text{Eq}$ for $\chi \in[0.3,0.6]$ (Fig. \ref{Figure1}B). The system at $\phi = 0.68$ also showed a reentrant behavior. However, at the two larger $\phi$ values, increasing $\chi$ from zero to 0.3 enhanced crystallinity, but a further increase did not melt the system (Fig. \ref{Figure1}C, see Supplementary Figs. 2, 3, and 4 for $|\psi^i_6|$ at other $\phi$ and $\chi$ values.). In fact, the strength of this reentrance, $\Delta = {\langle|\psi_6|\rangle_{\chi = 1}\over{\langle|\psi_6|\rangle_{\chi = 0.6}}}$, is non-monotonic with $\phi$, with a maximum at $\phi = 0.72$ (Fig. \ref{Figure1}D).

Since reentrant melting of the bulk is driven by chiral activity, the physics underlying the transition may be due entirely to odd effects, which can couple azimuthal flows to radial density changes \cite{bililign2022motile,fruchart2023odd,banerjee2017odd,mecke2023simultaneous}. The radial density, $\phi_A(r)$, shown in Fig. \ref{Figure1}E for $\phi = 0.72$ indeed showed marked changes with $\chi$. For $\chi = 0$ (net zero torque density), $\phi_A(r)$ is nearly identical to a passive disk packing at the same $\phi$, as expected. The enhancement in density near the boundary is simply due to layering \cite{huisman1997layering, van1988layering}. Strikingly, on increasing $\chi$, layering became more pronounced, and $\phi_A(r)$ in the bulk ($r/R < 0.7$) also showed clear changes. Both of these are clear signatures of additional radial stresses at play.

\begin{figure}[tbp]
\includegraphics[width=1\textwidth]{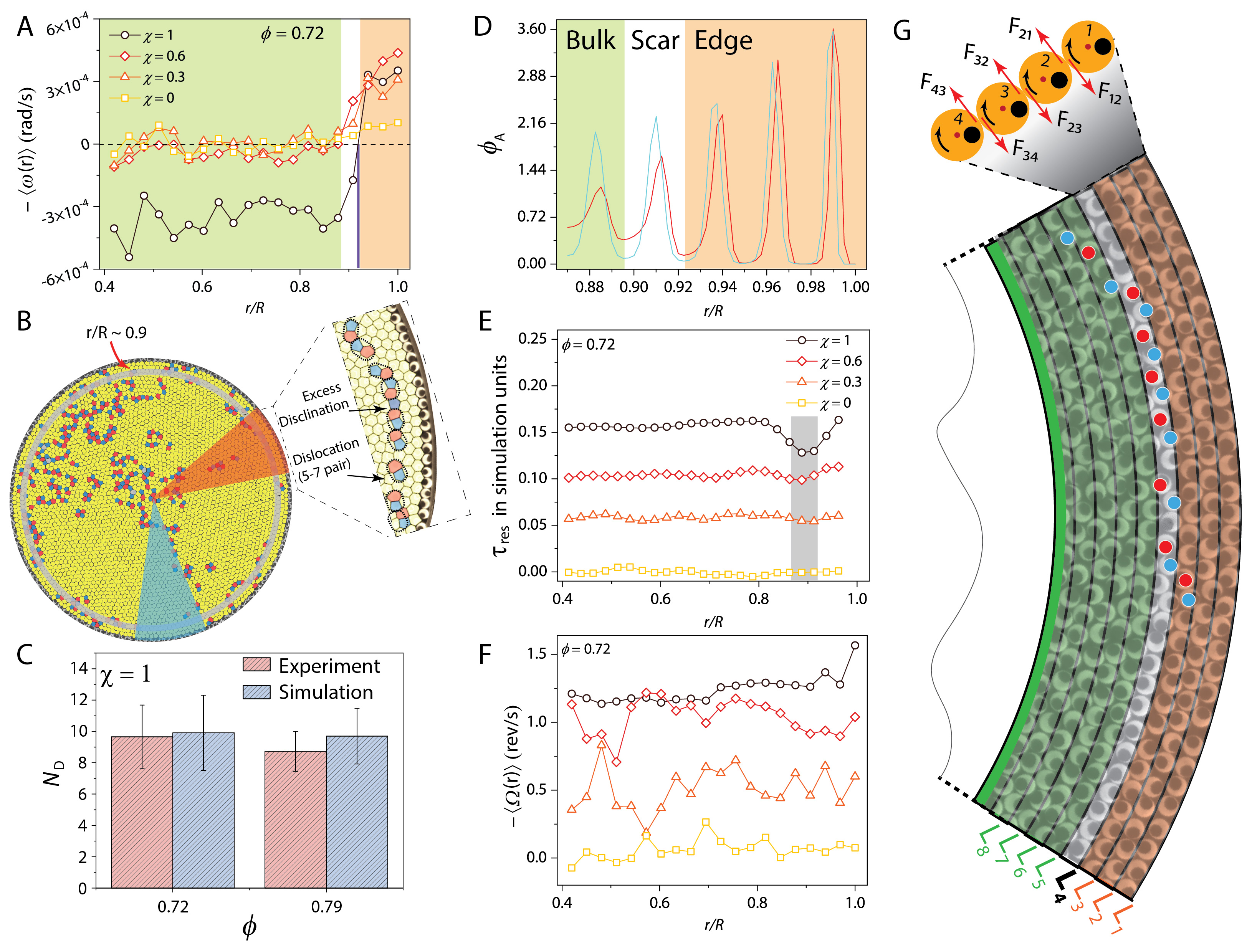}
\caption{\textbf{Grain boundary (GB) scars cause spinner crystals to self-shear.} All sub-figures are for $\phi = 0.72$. (\textbf{A}) Average angular velocity, $\omega(r)$, of spinners in annuli defined with respect to the system center versus $r/R$ for different values of $\chi$. There is a sharp drop in $\omega(r)$ at $r/R\approx 0.9$ (black vertical line) for $\chi>0$. The bulk and the edge rotate in opposite directions for $\chi = 0.6$ \& 1. The orange- and green-shaded regions correspond to the edge and bulk flow, respectively. (\textbf{B}) Voronoi tesselation of the spinner assembly at $\chi = 1$. The colors represent the spinner coordination number. The drop in $\omega(r)$ occurs across the annulus harboring GB scars (grey-shaded annulus). The zoomed-in view shows the composition of a GB scar. Unlike conventional grain boundaries, which terminate at the system edge, these scars terminate within the system \cite{bausch2003grain}. (\textbf{C}) The average number of dislocations per scar, $N_{\text{D}}$, for the experiment (red) and simulation (blue) at $\chi = 1$ for $\phi = 0.72$ and $\phi = 0.79$. The error bars represent the standard error. (\textbf{D}) shows $\phi_A(r)$ close to the edge for $\chi = 1$ in sectors with and without a GB scar, corresponding to the red- and blue-shaded regions in \textbf{(B)}, respectively. (\textbf{E}) Resistive torque, $\tau_{\text{res}}$, obtained from simulations for different values of $\chi$ at $\phi = 0.72$. There is a clear minimum in $\tau_{\text{res}}$ at $r/R\approx 0.9$ (grey-shaded region). (\textbf{F}) shows the annular spinner spin velocity $\langle\Omega(r)\rangle$ versus $r/R$ for different $\chi$ values. (\textbf{G}) A snapshot of spinners near the confining boundary at $\chi=1$. Blue and red dots are scattered over the five- and seven-coordinated spinners, respectively, to highlight the scar, which is located in layer $L_4$. The direction of the force experienced (exerted) by the annulus from (on) its neighboring one is also shown. For instance, $\bf{F}_{12}$ is the force on layer $L_1$ due to $L_2$. 
\justifying
}
\label{Figure2}
\end{figure}

These observations presented a natural segue into quantifying the flows that may have caused these density changes. In Fig. \ref{Figure2}A, we show annular angular velocity, $\omega(r)$, for different values of $\chi$ for $\phi=0.72$. For $\chi>0$, we observed edge flows - a ubiquitous feature of confined spinner assemblies. These flows are a consequence of the unbalanced torques on spinners at the edge \cite{van2016spatiotemporal, liu2020oscillating}. Additionally, the edge flows here have the same handedness as the spin of the preponderant spinner species ($\otimes$ spinners), indicating that inter-spinner frictional interactions dominate over spinner-boundary ones \cite{liu2020oscillating} (see Supplementary Fig. 5). The transverse forces exerted by spinners in the penultimate layer on those in the outermost one drive the edge flow. For $\chi = 0$, these forces cancel since $N^\otimes \approx N^\odot$ in each annulus, and $\omega(r)$ fluctuated around zero for all $r/R$.

Two features in the measured $\omega(r)$ stood out. First, the magnitude of the edge flow for $\chi>0$ dropped precipitously at $r/R \approx 0.9$ for $0.68\leq\phi\leq0.75$. This behavior is distinct from observations made hitherto in spinner liquids and solids. In the former, the edge flow decays exponentially into the bulk since the system lacks a shear modulus, while the finite modulus in the latter causes the flow to couple with the bulk, resulting in rigid body rotation \cite{van2016spatiotemporal}. Second, for $\phi = 0.68$ (Supplementary Figs. 6 and 18) and $0.72$, the system spontaneously \textit{self-shears}. Specifically, for $\phi = 0.72$, while $\omega(r)$ hovered around zero for $r/R<0.9$ at $\chi = 0.3$, it became negative and nearly constant in bulk for $\chi = 0.6$ and 1, with the magnitude of counterclockwise rotating bulk flow larger at $\chi = 1$ (see Supplementary Videos 1, 2, 3, and 4).

In contrast, at $\phi = 0.75$ and $\phi = 0.79$, where the system did not show reentrant melting with $\chi$, self-shearing was also absent (Supplementary Figs. 6 and 18). For $\phi = 0.75$ and at any given $\chi$, the magnitude of the edge current was smaller than that at $\phi = 0.72$ due to enhanced inter-spinner interactions, and $\omega(r)$ decayed to zero in the bulk, while for $\phi = 0.79$, we observed rigid body rotation for $\chi>0$. Evidently, reentrant melting and self-shearing go hand-in-hand, and this is the observation we will attempt to explain.

We first looked for cues in the structure of the spinner packings. In Fig. \ref{Figure2}B, we show the Voronoi tesselation of the spinner packing for $\phi = 0.72$ and $\chi = 1$. The yellow hexagons represent spinners with six-fold coordination (crystalline particles), while polygons of other colors represent under- and over-coordinated spinners (disclination defects). The large defect clusters in the bulk correspond to the liquid pocket shown in Fig. \ref{Figure1}B. Notably, at $r/R \approx 0.9$, where $\omega(r)$ fell sharply (grey annulus in Fig. \ref{Figure2}B), we observed azimuthally aligned grain boundary (GB) scars, which also appear as strings of low $|\psi_6|$ particles in Fig. \ref{Figure1}B. Each scar comprised of an excess 5-coordinated disclination decorated by a string of dislocations.  The dislocations - neutral disclination dipoles - are not a topological requirement but are present to help screen the strain field of the disclination \cite{bowick2000interacting, bowick2009two}. We could easily identify six GB scars when crystalline order was prominent ($\chi = 0.3 \text{ \& } 0.6$), while for $\chi = 0 \text{ \& } 1$, some of the scars were part of larger defect clusters because the bulk had melted, making their identification challenging (Supplementary Fig. 7). Previous studies on passive particle packings on a sphere observed that the number of excess dislocations per scar, $N_{D}$, depended solely on the ratio $D/a$ \cite{bowick2000interacting, bausch2003grain}. Besides $N_D$, even the scar configuration for passive disk packings at the same $\phi$ as the spinner packings were strikingly similar, demonstrating that it was the circular confinement, rather than the activity, that decided them. Notably, even our simulations of confined spinner packings - using values of $D$, $a$, and $\phi$ from experiments - closely matched our experimental results (Fig. \ref{Figure2}C, Supplementary Fig. 17). Although GB scars were not well-defined at $\phi = 0.68$ as the system is quite disordered, $\omega(r)$ dropped steeply in the annuli containing a higher-than-average defect density (see Supplementary Figs. 8 and 9).

GB scars/defects clearly impact the coupling between the edge flow and the bulk. To see why, we note that within a continuum description of a chiral active material with strains $\partial_lu_k$ and strain-rates $\partial_l \dot{u}_k$, the stress tensor $\sigma_{ij} = -P\delta_{ij}+K_{ijkl}\partial_lu_k+\eta_{ijkl}\partial_l \dot{u}_k+\sigma^{\text{spin}}_{ij}$ \cite{scheibner2020odd, fruchart2023odd, bililign2022motile}. Here, $P$ is the pressure, $\delta_{ij}$ is the Kronecker delta, and $K_{ijkl}$ and $\eta_{ijkl}$ are the elasticity and the viscosity tensors, respectively, and contain all moduli consistent with broken parity and time-reversal. The last term is the antisymmetric stress due to the imposed drive torque and is given by $\sigma^{\text{spin}}_{ij} = 2\eta_R\epsilon_{ij}\Omega$, where $\eta_R$ is the rotational viscosity, $\Omega$ is the spin angular velocity field of the particles and $\epsilon_{ij}$ is the Levi-Civita symbol \cite{tsai2005chiral}. A mismatch between $\Omega$ and the 2D vorticity, $(\nabla \times \mathbf{v})_z = 2\omega(r)$, where $\bf{v}$ is the velocity field, results in a frictional stress $\sigma^{\text{fric}}_{ij} = \eta_R\epsilon_{ij}(2\Omega-\omega)$ \cite{tsai2005chiral}. Since $\eta_R$ results from frictional collisions between the spinners, it is proportional to the local spinner density and is expressed as $\eta_R \propto\phi^2g(r_s)$ \cite{liu2020oscillating, han2021fluctuating}, where $g(r_s)$ is the pair-correlation function at contact. 

Here, in addition to layering, GB scars also alter the spinner density in the annuli, but only at $r/R\approx 0.9$. Fig. \ref{Figure2}D shows that $\phi_A(r/R\approx 0.9)$ in the sector harboring the scar is nearly 25\% smaller than the one without it for $\phi = 0.72$ (see Supplementary Fig. 10 for $\phi = 0.68$ and $\phi = 0.75$). Thus, the average spinner density in these annuli is smaller in a system with GB scars compared to one without, and consequently, $\eta_R$ and $\sigma^{\text{fric}}_{ij}$ should also be small for these annuli. A proxy for $\sigma^{\text{fric}}_{ij}$ that can be easily measured in our simulations is, $\tau_{\text{res}}(r) = \sum_{i \ne j} \tau_{ij}$ - the resistive torque experienced by an annulus from annuli straddling it. Here, $i$ and $j$ denote particles in the annulus under consideration and in those adjacent, respectively. Fig. \ref{Figure2}E shows $\tau_{\text{res}}(r)$ for the same $\chi$ values investigated in the experiments for $\phi = 0.72$. There is a pronounced minimum in $\tau_{\text{res}(r)}$ at $r/R\approx 0.9$ for $\chi = 1$ (see Supplementary Fig. 19 for other $\phi$ values), indicating that GB scars indeed weaken the coupling between the edge flow and the bulk. On lowering $\chi$, both the magnitude of $\tau_{\text{res}}(r)$ and also the depth of the minimum become smaller and vanish for $\chi = 0$. This is because increasing the fraction of $\odot$ spinners at the expense of $\otimes$ spinners systematically reduces annulus-averaged spin angular velocity, $\langle\Omega(r)\rangle$, and consequently, $\tau_{\text{res}(r)}$ (see Fig. \ref{Figure2}F) (see Supplementary Fig. 12 for other values of $\phi$ and $\chi$).

We can now explain why the system self-shears. We note that while the spinners in the edge layers (labeled $L_1-L_3$ in Fig. \ref{Figure2}G) are mostly in the registry and, hence, move as a plug, GB scars in layer $L_4$ disrupt layering and prevent this registry from extending into the bulk. The clockwise rotation of spinners in layer $L_3$ exerts a counterclockwise tangential force on layer $L_4$, and because, $\phi_A^{\text{L3}} > \phi_A^{\text{L4}}$ (Fig. \ref{Figure1}D), layer $L_4$ rotates counterclockwise. spinners in layers further interior fall in the registry again as the scars have eased the frustration caused by circular confinement. The counterclockwise rotation of layer $L_4$ is now coupled to the bulk, and the system spontaneously self-shears. On decreasing $\chi$, however, the magnitude of the counterclockwise tangential force exerted by $L_3$ on $L_4$ also decreases, and self-shearing weakens.

\begin{figure}[tbp]
\includegraphics[width=1\textwidth]{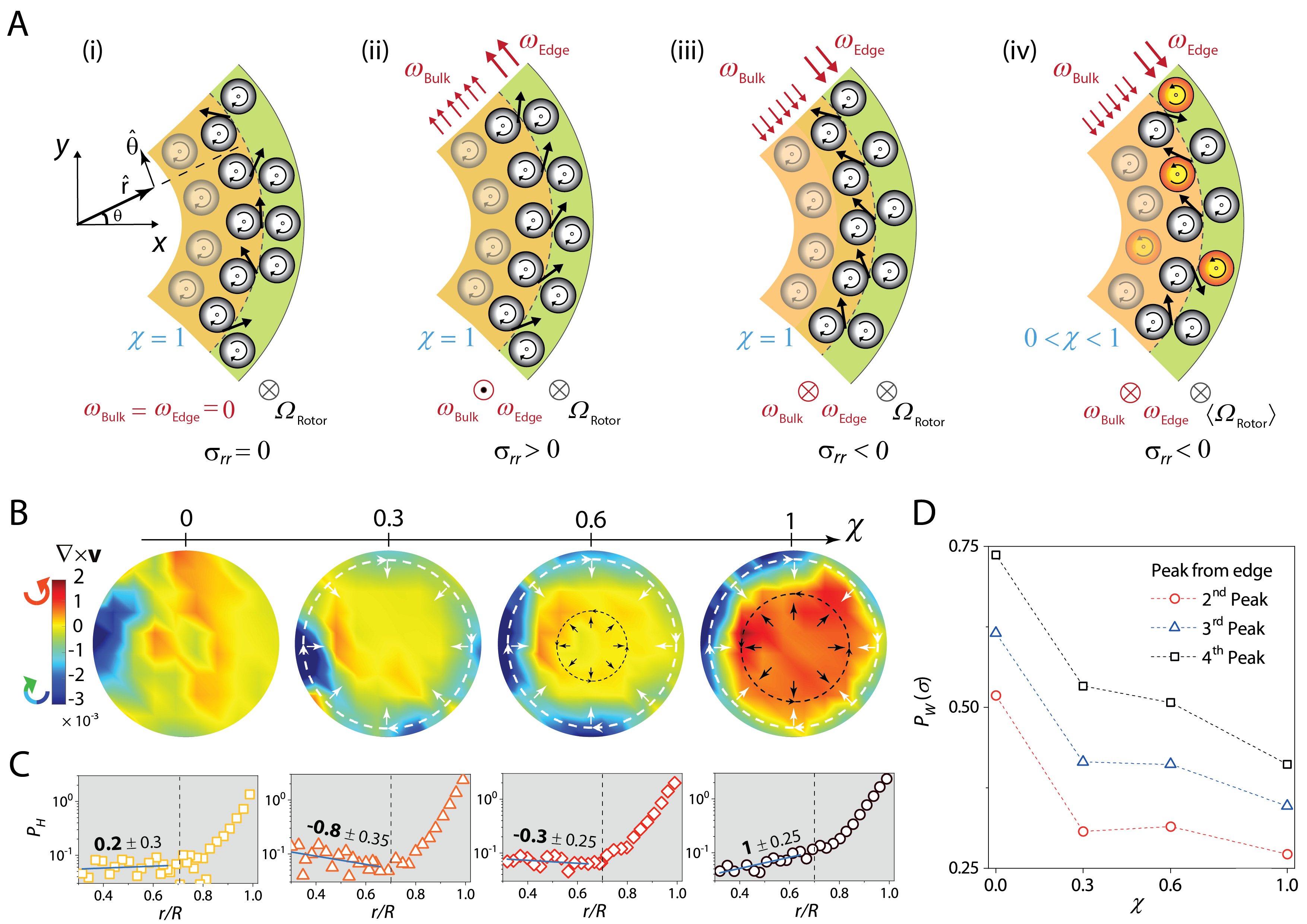}
\caption{\textbf{An odd re-entrant melting transition.} (\textbf{A}) Illustration of the microscopic mechanism that gives rise to odd radial stresses in spinner materials. The handedness of particle spin, edge, and bulk flows is represented by $(\odot)$ for counterclockwise and $(\otimes)$ for a clockwise motion. Panels (i)-(iii) are for $\otimes$ spinners. Panel (i): No flow ($\omega_\text{Edge} = \omega_\text{Bulk} = 0$). Here, edge spinners collide with bulk spinners from above and below with equal probability. During collisions, the transverse forces (black arrows) on average add up to zero along $\hat{r}$ and hence the radial stress $\sigma_{rr}=0$. Panel (ii): Edge and bulk flow have handedness opposite to particle spin. Edge spinners collide with bulk spinners more frequently from below since $^\odot \omega_\text{Edge}>^\odot \omega_\text{Bulk}$. The transverse forces now have a component along $\hat{r}$, and, hence, $\sigma_{rr}>0$. The bulk dilates. Panel (iii): Edge and bulk flow and particle spin have the same handedness and $^\otimes \omega_\text{Edge}>^\otimes \omega_\text{Bulk}$. $\sigma_{rr}$ now points inwards, compressing the bulk. Panel (iv): For $0<\chi<1$, although $^\otimes \omega_\text{Edge}>^\otimes \omega_\text{Bulk}$, transverse forces in many inter-spinner collisions have a component along $\hat{r}$ unlike in Panel (iii) where it is predominantly along $-\hat{r}$. Thus, the stress $|\sigma'_{rr}|$ is always smaller than the $\chi = 1$ case. $\sigma'_{rr} = 0$ for $\chi = 0$. \textbf{(B)}, \textbf{(C)}, and \textbf{(D)} are for $\phi = 0.72$. \textbf{(B)} shows the vorticity obtained from a coarse-grained velocity field for different $\chi$ values. The color bar denotes the magnitude and handedness of the vorticity. The black and the white arrows correspond to $\sigma_{rr}$ along $\hat{r}$ and $-\hat{r}$, respectively. (\textbf{C}) shows the peak height, $P_H(r)$, of $\phi_A(r)$ for different $\chi$ values. The peak height is measured from the baseline shown in Fig. \ref{Figure1}D. The dashed vertical lines delineate the bulk from the edge. (\textbf{D}) shows the full width at half maximum (FWHM) in units of $d_P$ for peaks close to the edge. The peak width is a direct measure of the total $\sigma_{rr}$.
\justifying
}
\label{Figure3}   
\end{figure}

Having uncovered why tuning $\chi$ produces the observed flows, we now show that in the presence of these flows, parity-violating inter-spinner collisions generate radial odd stress, $\sigma_{rr}$, which drives reentrant melting. Essentially, due to transverse inter-spinner interactions, edge and bulk flows having handedness identical (opposite) to the particle spin produce a radially inward (outward) $\sigma_{rr}$ (Fig. \ref{Figure3}A) \cite{yang2021topologically, banerjee2017odd, mecke2023simultaneous}. This inward stress compresses the bulk, while the outward stress dilates it. Importantly, $\sigma_{rr}$ breaks parity and is linked with odd material moduli because changes in spinner density are transverse to the flow (see Supplementary Fig. 16).

The coarse-grained local vorticity, presented in Fig. \ref{Figure3}B for all values of $\chi$ at $\phi = 0.72$, provides a precise representation of the flow field in our system. We calculated vorticity solely for the longest-lived flows since only these significantly influence $\phi_A(r)$. As a readout for radial compression and dilation, we estimated the height of the peaks, $P_H$, of $\phi_A(r)$ (shown in Fig. \ref{Figure1}D) and plotted it as a function of $r/R$ for different values of $\chi$ (Fig. \ref{Figure3}C). A negative (positive) slope of $P_H(r)$ for $r/R < 0.7$ indicates a compressed (dilated) bulk. For $\chi = 0$, we observed small and weak $\otimes$ and $\odot$ vortices that had no impact on the bulk spinner density - $P_H(r)$ is constant for $r/R\leq 0.7$; the peak widths are at their largest (Fig. \ref{Figure3}D). Because layering at the wall depletes the bulk, $\phi_\text{Bulk}\approx 0.706<\phi^\text{Liq}_\text{Eq}$, and the assembly is liquid-like. As we increase $\chi$ to 0.3, the $\otimes$ edge flow results in a $-\sigma_{rr}\hat{r}$, which compresses the bulk and causes the system to crystallize ($\phi_\text{Bulk} \approx 0.725>\phi^\text{Xtal}_\text{Eq}$); $|\psi_6|$ is large (Fig. \ref{Figure1}C). The peak widths are also smaller compared to $\chi = 0$. For $\chi = 0.6$ and 1, since the system self-shears, the enhancement in $\phi_\text{Bulk}$ due to an inward radial stress from the $\otimes$ edge flow is counteracted by an outward stress from the $\odot$ bulk flow; the peak width is reduced further (Fig. \ref{Figure3}D). The contribution from the edge barely wins over and compresses the bulk at $\chi=0.6$, but this suffices to crystallize the bulk. At $\chi=1$, the outward radial stress from the bulk overwhelms the inward stress from the edge; the peak widths are at their smallest. The bulk dilates $\phi_\text{Bulk}\approx\phi^\text{Liq}_\text{Eq}$, and the system melts again. This striking link between odd radial stress and bulk density changes is present at all $\phi$ and $\chi$ values studied here (see Supplementary Figs. 13, 14, and 15).

In summary, we have discovered that grain boundary scars dramatically alter the behavior of confined active spinner crystals. These scars function as weak links and enable the topologically-protected edge flow to detach from the bulk, triggering spontaneous self-shearing with increasing chiral activity. Parity-violating odd stresses couple the shear flows to density changes, driving the observed reentrant melting transition. The observation that the strength of reentrant melting is maximum at a density near the equilibrium melting density (Fig. \ref{Figure1}D) may not be incidental since, in this regime, even small odd stresses can trigger significant phase changes. Through the strategic design of the confining boundary geometry, we can now control the type and spatial arrangement of defects/scars, offering a powerful tool to engineer flows and induce novel functionalities in these odd materials. Our findings pave the way for an exciting future where the intricate interplay of odd elasticity and geometrical frustration unveils physics that is increasingly odd.

\newpage
\section{Author Contributions}
\textbf{U.T.} and \textbf{R.G.} planned and executed experimental research and designed and performed analysis. \textbf{R.M.} performed simulations. \textbf{P.A.} contributed to project design and trained \textbf{U.T.} on experimental methodology. \textbf{A.K.S.} and \textbf{S.R.} contributed to project development. \textbf{U.T.}, \textbf{R.M.} and \textbf{R.G.} wrote the paper with inputs from all authors.

\section{Acknowledgments}
\textbf{U.T.} thanks the Jawaharlal Nehru Centre for Advanced Scientific Research, Bangalore, India, for a research fellowship. \textbf{P.A.} thanks the Jawaharlal Nehru Centre for Advanced Scientific Research, Bangalore, India, for a research fellowship. \textbf{A.K.S.} thanks the Science and Engineering Research Board, Government of India, for the National Science Chair. \textbf{S.R.} acknowledges a J C Bose Fellowship of the ANRF, INDIA. \textbf{R.G.} thanks the Department of Science and Technology, India, for financial support through the Swarnajayanti Fellowship Grant (DST/SJF/PSA-03/2017-22). \textbf{U.T.} and \textbf{R.G.} thank Shreyas Gokhale for critical inputs on the manuscript.

\bibliographystyle{ieeetr}
\bibliography{references}

@book{nelson2002defects,
  title={Defects and geometry in condensed matter physics},
  author={Nelson, David R},
  year={2002},
  publisher={Cambridge University Press}
}

@book{chaikin1995principles,
  title={Principles of condensed matter physics},
  author={Chaikin, Paul M and Lubensky, Tom C and Witten, Thomas A},
  volume={10},
  year={1995},
  publisher={Cambridge university press Cambridge}
}

@article{jangizehi2020defects,
  title={Defects and defect engineering in Soft Matter},
  author={Jangizehi, Amir and Schmid, Friederike and Besenius, Pol and Kremer, Kurt and Seiffert, Sebastian},
  journal={Soft Matter},
  volume={16},
  number={48},
  pages={10809--10859},
  year={2020},
  publisher={Royal Society of Chemistry}
}

@article{shankar2022topological,
  title={Topological active matter},
  author={Shankar, Suraj and Souslov, Anton and Bowick, Mark J and Marchetti, M Cristina and Vitelli, Vincenzo},
  journal={Nature Reviews Physics},
  volume={4},
  number={6},
  pages={380--398},
  year={2022},
  publisher={Nature Publishing Group UK London}
}

@article{ardavseva2022topological,
  title={Topological defects in biological matter},
  author={Arda{\v{s}}eva, Aleksandra and Doostmohammadi, Amin},
  journal={Nature Reviews Physics},
  volume={4},
  number={6},
  pages={354--356},
  year={2022},
  publisher={Nature Publishing Group UK London}
}

@book{sadoc1999geometrical,
  title={Geometrical frustration},
  author={Sadoc, Jean-Fran{\c{c}}ois and Mosseri, Rimy},
  year={1999},
  publisher={Cambridge University Press}
}

@article{bowick2009two,
  title={Two-dimensional matter: order, curvature and defects},
  author={Bowick, Mark J and Giomi, Luca},
  journal={Advances in Physics},
  volume={58},
  number={5},
  pages={449--563},
  year={2009},
  publisher={Taylor \& Francis}
}

@article{bruinsma2021physics,
  title={Physics of viral dynamics},
  author={Bruinsma, Robijn F and Wuite, Gijs JL and Roos, Wouter H},
  journal={Nature Reviews Physics},
  volume={3},
  number={2},
  pages={76--91},
  year={2021},
  publisher={Nature Publishing Group UK London}
}

@article{bausch2003grain,
  title={Grain boundary scars and spherical crystallography},
  author={Bausch, AR and Bowick, Mark John and Cacciuto, A and Dinsmore, AD and Hsu, MF and Nelson, DR and Nikolaides, MG and Travesset, A and Weitz, DA},
  journal={Science},
  volume={299},
  number={5613},
  pages={1716--1718},
  year={2003},
  publisher={American Association for the Advancement of Science}
}

@article{guerra2018freezing,
  title={Freezing on a sphere},
  author={Guerra, Rodrigo E and Kelleher, Colm P and Hollingsworth, Andrew D and Chaikin, Paul M},
  journal={Nature},
  volume={554},
  number={7692},
  pages={346--350},
  year={2018},
  publisher={Nature Publishing Group UK London}
}

@article{singh2022observation,
  title={Observation of two-step melting on a sphere},
  author={Singh, Navneet and Sood, AK and Ganapathy, Rajesh},
  journal={Proceedings of the National Academy of Sciences},
  volume={119},
  number={32},
  pages={e2206470119},
  year={2022},
  publisher={National Academy of Sciences}
}

@article{moore1999absence,
  title={Absence of a finite-temperature melting transition in the classical two-dimensional one-component plasma},
  author={Moore, MA and P{\'e}rez--Garrido, A},
  journal={Physical Review Letters},
  volume={82},
  number={20},
  pages={4078},
  year={1999},
  publisher={APS}
}

@article{amore2023thomson,
  title={Thomson problem in the disk},
  author={Amore, Paolo and Zarate, Ulises},
  journal={Physical Review E},
  volume={108},
  number={5},
  pages={055302},
  year={2023},
  publisher={APS}
}

@article{bowick2000interacting,
  title={Interacting topological defects on frozen topographies},
  author={Bowick, Mark J and Nelson, David R and Travesset, Alex},
  journal={Physical Review B},
  volume={62},
  number={13},
  pages={8738},
  year={2000},
  publisher={APS}
}

@article{lipowsky2005direct,
  title={Direct visualization of dislocation dynamics in grain-boundary scars},
  author={Lipowsky, Peter and Bowick, Mark J and Meinke, Jan H and Nelson, David R and Bausch, Andreas R},
  journal={Nature Materials},
  volume={4},
  number={5},
  pages={407--411},
  year={2005},
  publisher={Nature Publishing Group UK London}
}

@article{negri2015deformation,
  title={Deformation and failure of curved colloidal crystal shells},
  author={Negri, Carlotta and Sellerio, Alessandro L and Zapperi, Stefano and Miguel, M Carmen},
  journal={Proceedings of the National Academy of Sciences},
  volume={112},
  number={47},
  pages={14545--14550},
  year={2015},
  publisher={National Academy of Sciences}
}

@article{marchetti2013hydrodynamics,
  title={Hydrodynamics of soft active matter},
  author={Marchetti, M Cristina and Joanny, Jean-Fran{\c{c}}ois and Ramaswamy, Sriram and Liverpool, Tanniemola B and Prost, Jacques and Rao, Madan and Simha, R Aditi},
  journal={Reviews of Modern Physics},
  volume={85},
  number={3},
  pages={1143--1189},
  year={2013},
  publisher={APS}
}

@article{narayan2007long,
  title={Long-lived giant number fluctuations in a swarming granular nematic},
  author={Narayan, Vijay and Ramaswamy, Sriram and Menon, Narayanan},
  journal={Science},
  volume={317},
  number={5834},
  pages={105--108},
  year={2007},
  publisher={American Association for the Advancement of Science}
}

@article{giomi2014defect,
  title={Defect dynamics in active nematics},
  author={Giomi, Luca and Bowick, Mark J and Mishra, Prashant and Sknepnek, Rastko and Cristina Marchetti, M},
  journal={Philosophical Transactions of the Royal Society A: Mathematical, Physical and Engineering Sciences},
  volume={372},
  number={2029},
  pages={20130365},
  year={2014},
  publisher={The Royal Society Publishing}
}

@article{grossmann2020particle,
  title={A particle-field approach bridges phase separation and collective motion in active matter},
  author={Gro{\ss}mann, Robert and Aranson, Igor S and Peruani, Fernando},
  journal={Nature Communications},
  volume={11},
  number={1},
  pages={5365},
  year={2020},
  publisher={Nature Publishing Group UK London}
}

@article{arora2022motile,
  title={Motile topological defects hinder dynamical arrest in dense liquids of active ellipsoids},
  author={Arora, Pragya and Sood, AK and Ganapathy, Rajesh},
  journal={Physical Review Letters},
  volume={128},
  number={17},
  pages={178002},
  year={2022},
  publisher={APS}
}

@article{wu2017transition,
  title={Transition from turbulent to coherent flows in confined three-dimensional active fluids},
  author={Wu, Kun-Ta and Hishamunda, Jean Bernard and Chen, Daniel TN and DeCamp, Stephen J and Chang, Ya-Wen and Fern{\'a}ndez-Nieves, Alberto and Fraden, Seth and Dogic, Zvonimir},
  journal={Science},
  volume={355},
  number={6331},
  pages={eaal1979},
  year={2017},
  publisher={American Association for the Advancement of Science}
}

@article{memarian2024controlling,
  title={Controlling chaos: Periodic defect braiding in active nematics confined to a cardioid},
  author={Memarian, Fereshteh L and Hammar, Derek and Sabbir, Md Mainul Hasan and Elias, Mark and Mitchell, Kevin A and Hirst, Linda S},
  journal={Physical Review Letters},
  volume={132},
  number={22},
  pages={228301},
  year={2024},
  publisher={APS}
}

@article{liu2021viscoelastic,
  title={Viscoelastic control of spatiotemporal order in bacterial active matter},
  author={Liu, Song and Shankar, Suraj and Marchetti, M Cristina and Wu, Yilin},
  journal={Nature},
  volume={590},
  number={7844},
  pages={80--84},
  year={2021},
  publisher={Nature Publishing Group UK London}
}

@article{williams2022confinement,
  title={Confinement-induced accumulation and de-mixing of microscopic active-passive mixtures},
  author={Williams, Stephen and Jeanneret, Rapha{\"e}l and Tuval, Idan and Polin, Marco},
  journal={Nature Communications},
  volume={13},
  number={1},
  pages={4776},
  year={2022},
  publisher={Nature Publishing Group UK London}
}

@article{theillard2017geometric,
  title={Geometric control of active collective motion},
  author={Theillard, Maxime and Alonso-Matilla, Roberto and Saintillan, David},
  journal={Soft Matter},
  volume={13},
  number={2},
  pages={363--375},
  year={2017},
  publisher={Royal Society of Chemistry}
}

@article{negi2023geometry,
  title={Geometry-induced dynamics of confined chiral active matter},
  author={Negi, Archit and Beppu, Kazusa and Maeda, Yusuke T},
  journal={Physical Review Research},
  volume={5},
  number={2},
  pages={023196},
  year={2023},
  publisher={APS}
}

@article{hardouin2019reconfigurable,
  title={Reconfigurable flows and defect landscape of confined active nematics},
  author={Hardo{\"u}in, J{\'e}r{\^o}me and Hughes, Rian and Doostmohammadi, Amin and Laurent, Justine and Lopez-Leon, Teresa and Yeomans, Julia M and Ign{\'e}s-Mullol, Jordi and Sagu{\'e}s, Francesc},
  journal={Communications Physics},
  volume={2},
  number={1},
  pages={121},
  year={2019},
  publisher={Nature Publishing Group UK London}
}

@article{peng2016command,
  title={Command of active matter by topological defects and patterns},
  author={Peng, Chenhui and Turiv, Taras and Guo, Yubing and Wei, Qi-Huo and Lavrentovich, Oleg D},
  journal={Science},
  volume={354},
  number={6314},
  pages={882--885},
  year={2016},
  publisher={American Association for the Advancement of Science}
}

@article{zhang2021spatiotemporal,
  title={Spatiotemporal control of liquid crystal structure and dynamics through activity patterning},
  author={Zhang, Rui and Redford, Steven A and Ruijgrok, Paul V and Kumar, Nitin and Mozaffari, Ali and Zemsky, Sasha and Dinner, Aaron R and Vitelli, Vincenzo and Bryant, Zev and Gardel, Margaret L and others},
  journal={Nature Materials},
  volume={20},
  number={6},
  pages={875--882},
  year={2021},
  publisher={Nature Publishing Group UK London}
}

@article{gompper20202020,
  title={The 2020 motile active matter roadmap},
  author={Gompper, Gerhard and Winkler, Roland G and Speck, Thomas and Solon, Alexandre and Nardini, Cesare and Peruani, Fernando and L{\"o}wen, Hartmut and Golestanian, Ramin and Kaupp, U Benjamin and Alvarez, Luis and others},
  journal={Journal of Physics: Condensed Matter},
  volume={32},
  number={19},
  pages={193001},
  year={2020},
  publisher={IOP Publishing}
}

@article{ramaswamy2010mechanics,
  title={The mechanics and statistics of active matter},
  author={Ramaswamy, Sriram},
  journal={Annu. Rev. Condens. Matter Phys.},
  volume={1},
  number={1},
  pages={323--345},
  year={2010},
  publisher={Annual Reviews}
}

@article{keber2014topology,
  title={Topology and dynamics of active nematic vesicles},
  author={Keber, Felix C and Loiseau, Etienne and Sanchez, Tim and DeCamp, Stephen J and Giomi, Luca and Bowick, Mark J and Marchetti, M Cristina and Dogic, Zvonimir and Bausch, Andreas R},
  journal={Science},
  volume={345},
  number={6201},
  pages={1135--1139},
  year={2014},
  publisher={American Association for the Advancement of Science}
}

@article{sknepnek2015active,
  title={Active swarms on a sphere},
  author={Sknepnek, Rastko and Henkes, Silke},
  journal={Physical Review E},
  volume={91},
  number={2},
  pages={022306},
  year={2015},
  publisher={APS}
}

@article{wioland2013confinement,
  title={Confinement stabilizes a bacterial suspension into a spiral vortex},
  author={Wioland, Hugo and Woodhouse, Francis G and Dunkel, J{\"o}rn and Kessler, John O and Goldstein, Raymond E},
  journal={Physical Review Letters},
  volume={110},
  number={26},
  pages={268102},
  year={2013},
  publisher={APS}
}

@article{fruchart2023odd,
  title={Odd viscosity and odd elasticity},
  author={Fruchart, Michel and Scheibner, Colin and Vitelli, Vincenzo},
  journal={Annual Review of Condensed Matter Physics},
  volume={14},
  number={1},
  pages={471--510},
  year={2023},
  publisher={Annual Reviews}
}

@article{tsai2005chiral,
  title={A chiral granular gas},
  author={Tsai, J-C and Ye, Fangfu and Rodriguez, Juan and Gollub, Jerry P and Lubensky, TC},
  journal={Physical Review Letters},
  volume={94},
  number={21},
  pages={214301},
  year={2005},
  publisher={APS}
}

@article{banerjee2017odd,
  title={Odd viscosity in chiral active fluids},
  author={Banerjee, Debarghya and Souslov, Anton and Abanov, Alexander G and Vitelli, Vincenzo},
  journal={Nature Communications},
  volume={8},
  number={1},
  pages={1573},
  year={2017},
  publisher={Nature Publishing Group UK London}
}

@article{mecke2023simultaneous,
  title={Simultaneous emergence of active turbulence and odd viscosity in a colloidal chiral active system},
  author={Mecke, Joscha and Gao, Yongxiang and Ram{\'\i}rez Medina, Carlos A and Aarts, Dirk GAL and Gompper, Gerhard and Ripoll, Marisol},
  journal={Communications Physics},
  volume={6},
  number={1},
  pages={324},
  year={2023},
  publisher={Nature Publishing Group UK London}
}

@article{yang2021topologically,
  title={Topologically protected transport of cargo in a chiral active fluid aided by odd-viscosity-enhanced depletion interactions},
  author={Yang, Qing and Zhu, Hongwei and Liu, Peng and Liu, Rui and Shi, Qingfan and Chen, Ke and Zheng, Ning and Ye, Fangfu and Yang, Mingcheng},
  journal={Physical Review Letters},
  volume={126},
  number={19},
  pages={198001},
  year={2021},
  publisher={APS}
}

@article{tan2022odd,
  title={Odd dynamics of living chiral crystals},
  author={Tan, Tzer Han and Mietke, Alexander and Li, Junang and Chen, Yuchao and Higinbotham, Hugh and Foster, Peter J and Gokhale, Shreyas and Dunkel, J{\"o}rn and Fakhri, Nikta},
  journal={Nature},
  volume={607},
  number={7918},
  pages={287--293},
  year={2022},
  publisher={Nature Publishing Group UK London}
}

@article{bililign2022motile,
  title={Motile dislocations knead odd crystals into whorls},
  author={Bililign, Ephraim S and Balboa Usabiaga, Florencio and Ganan, Yehuda A and Poncet, Alexis and Soni, Vishal and Magkiriadou, Sofia and Shelley, Michael J and Bartolo, Denis and Irvine, William TM},
  journal={Nature Physics},
  volume={18},
  number={2},
  pages={212--218},
  year={2022},
  publisher={Nature Publishing Group UK London}
}

@article{van2016spatiotemporal,
  title={Spatiotemporal order and emergent edge currents in active spinner materials},
  author={Van Zuiden, Benjamin C and Paulose, Jayson and Irvine, William TM and Bartolo, Denis and Vitelli, Vincenzo},
  journal={Proceedings of the National Academy of Sciences},
  volume={113},
  number={46},
  pages={12919--12924},
  year={2016},
  publisher={National Acad Sciences}
}

@article{liu2020oscillating,
  title={Oscillating collective motion of active rotors in confinement},
  author={Liu, Peng and Zhu, Hongwei and Zeng, Ying and Du, Guangle and Ning, Luhui and Wang, Dunyou and Chen, Ke and Lu, Ying and Zheng, Ning and Ye, Fangfu and others},
  journal={Proceedings of the National Academy of Sciences},
  volume={117},
  number={22},
  pages={11901--11907},
  year={2020},
  publisher={National Acad Sciences}
}

@article{yang2020robust,
  title={Robust boundary flow in chiral active fluid},
  author={Yang, Xiang and Ren, Chenyang and Cheng, Kangjun and Zhang, HP},
  journal={Physical Review E},
  volume={101},
  number={2},
  pages={022603},
  year={2020},
  publisher={APS}
}

@article{scholz2018rotating,
  title={Rotating robots move collectively and self-organize},
  author={Scholz, Christian and Engel, Michael and P{\"o}schel, Thorsten},
  journal={Nature Communications},
  volume={9},
  number={1},
  pages={931},
  year={2018},
  publisher={Nature Publishing Group UK London}
}

@article{petroff2023density,
  title={Density-mediated spin correlations drive edge-to-bulk flow transition in active chiral matter},
  author={Petroff, Alexander P and Whittington, Christopher and Kudrolli, Arshad},
  journal={Physical Review E},
  volume={108},
  number={1},
  pages={014609},
  year={2023},
  publisher={APS}
}

@article{bernard2011two,
  title={Two-step melting in two dimensions: first-order liquid-hexatic transition},
  author={Bernard, Etienne P and Krauth, Werner},
  journal={Physical Review Letters},
  volume={107},
  number={15},
  pages={155704},
  year={2011},
  publisher={APS}
}

@article{huisman1997layering,
  title={Layering of a liquid metal in contact with a hard wall},
  author={Huisman, Willem Jan and Peters, Joost F and Zwanenburg, Michel J and de Vries, Steven A and Derry, Trevor E and Abernathy, Douglas and van der Veen, J Friso},
  journal={Nature},
  volume={390},
  number={6658},
  pages={379--381},
  year={1997},
  publisher={Nature Publishing Group UK London}
}

@article{van1988layering,
  title={Layering in colloidal fluids near a smooth repulsive wall},
  author={Van Winkle, David H and Murray, CA},
  journal={The Journal of Chemical Physics},
  volume={89},
  number={6},
  pages={3885--3891},
  year={1988},
  publisher={American Institute of Physics}
}

@article{scheibner2020odd,
  title={Odd elasticity},
  author={Scheibner, Colin and Souslov, Anton and Banerjee, Debarghya and Sur{\'o}wka, Piotr and Irvine, William TM and Vitelli, Vincenzo},
  journal={Nature Physics},
  volume={16},
  number={4},
  pages={475--480},
  year={2020},
  publisher={Nature Publishing Group UK London}
}

@article{han2021fluctuating,
  title={Fluctuating hydrodynamics of chiral active fluids},
  author={Han, Ming and Fruchart, Michel and Scheibner, Colin and Vaikuntanathan, Suriyanarayanan and De Pablo, Juan J and Vitelli, Vincenzo},
  journal={Nature Physics},
  volume={17},
  number={11},
  pages={1260--1269},
  year={2021},
  publisher={Nature Publishing Group UK London}
}

@article{kole2024chirality,
  title={Chirality and odd mechanics in active columnar phases},
  author={Kole, SJ and Alexander, Gareth P and Maitra, Ananyo and Ramaswamy, Sriram},
  journal={PNAS nexus},
  volume={3},
  number={10},
  pages={pgae398},
  year={2024},
  publisher={Oxford University Press US}
}

@article{kole2021layered,
  title={Layered chiral active matter: Beyond odd elasticity},
  author={Kole, SJ and Alexander, Gareth P and Ramaswamy, Sriram and Maitra, Ananyo},
  journal={Physical Review Letters},
  volume={126},
  number={24},
  pages={248001},
  year={2021},
  publisher={APS}
}

@article{huang2023odd,
  title={Odd elasticity in driven granular matter},
  author={Huang, Rosalind and Mandal, Rituparno and Scheibner, Colin and Vitelli, Vincenzo},
  journal={arXiv preprint arXiv:2311.18720},
  year={2023}
}

\clearpage

\end{document}